\documentclass[12pt,a4paper]{article}
\pdfoutput=1
\usepackage{jheppub}
\usepackage[english]{babel}
\usepackage[utf8]{inputenc}
\hypersetup{unicode}

\makeatletter
\def\@fpheader{\relax}
\makeatother

\newcommand{\eq}{\begin{equation}}
\newcommand{\feq}{\end{equation}}

\newcommand{\RR}{\mathbb{R}}
\newcommand{\SL}{\mathrm{SL}}

\DeclareMathOperator{\diag}{diag}

\title{Rotating black holes with Nil or SL(2,\,$\mathbb{R}$) horizons}

\author[1,2]{Federico Faedo,}
\author[3,4]{Silke Klemm}
\author[4]{and Pietro Mariotti}

\affiliation[1]{Dipartimento di Matematica, Universit\`a di Torino, \\
Via Carlo Alberto 10, I-10123 Torino, Italy}
\affiliation[2]{INFN, Sezione di Torino, \\
Via Giuria 1, I-10125 Torino, Italy}
\affiliation[3]{INFN, Sezione di Milano, \\
Via Celoria 16, I-20133 Milano, Italy}
\affiliation[4]{Dipartimento di Fisica, Universit\`a di Milano, \\
Via Celoria 16, I-20133 Milano, Italy}

\emailAdd{federicomichele.faedo@unito.it}
\emailAdd{silke.klemm@mi.infn.it}
\emailAdd{pietro.mariotti@studenti.unimi.it}
\preprint{IFUM-1104-FT}

\abstract{We construct rotating black holes in $N=2$, $D=5$ minimal and matter-coupled gauged supergravity, with horizons that are homogeneous but not isotropic. Such spaces belong to the eight Thurston model geometries, out of which we consider the cases Nil and SL$(2,\mathbb{R})$.
In the former, we use the recipe of \cite{Gauntlett:2003fk} to directly rederive the solution that was obtained by Gutowski and Reall in \cite{Gutowski:2004ez} as a scaling limit from a spherical black hole.
With the same techniques, the first example of a black hole with SL$(2,\mathbb{R})$ horizon is constructed, which is rotating and one quarter BPS. The physical properties of this solution are discussed, and it is shown that in the near-horizon limit it boils down to the geometry of \cite{Gutowski:2004ez}, with a supersymmetry enhancement to one half. Dimensional reduction to $D=4$ gives a new solution with hyperbolic horizon to the t$^3$ model that carries both electric and magnetic charges.
Moreover, we show how to get a nonextremal rotating Nil black hole by applying a certain scaling limit to Kerr-AdS$_5$ with two equal rotation parameters, which consists in zooming onto the north pole of the S$^2$ over which the S$^3$ is fibered, while boosting the horizon velocity effectively to the speed of light.}


\begin{document}

\maketitle

\flushbottom

\section{Introduction}

In the seventies of the last century Hawking proved his famous theorem \cite{Hawking:1971vc,
Hawking:1973uf} on the topology of black holes, which asserts that event horizon cross-sections
of four-dimensional asymptotically flat stationary black holes obeying the dominant energy condition are
topologically $\text{S}^2$. This result extends to outer apparent horizons in black hole spacetimes that
are not necessarily stationary \cite{Hawking:72}. Such restrictive uniqueness theorems do not hold in
higher dimensions, the most famous counterexample being the black ring of Emparan and
Reall \cite{Emparan:2001wn}, with horizon topology $\text{S}^2\times\text{S}^1$.
Nevertheless, Galloway and Schoen \cite{Galloway:2005mf} were able to show that, in arbitrary
dimension, cross-sections of the event horizon (in the stationary case) and outer apparent horizons
(in the general case) are of positive Yamabe type, i.e., admit metrics of positive scalar curvature.

Instead of increasing the number of dimensions, one can relax some of the assumptions that go
into Hawking's theorem in order to have black holes with nonspherical topology. One such possibility
is to add a negative cosmological constant~$\Lambda$. Interpreting the term $-\Lambda g_{\mu\nu}$
as $8\pi G$ times the energy-momentum tensor $T_{\mu\nu}$, one has obviously that
$-T_{\mu\nu}\xi^\nu$ is past-pointing for every future-pointing causal vector~$\xi^\nu$, and thus
a violation of the dominant energy condition. Moreover, since for $\Lambda<0$ the solutions generically
asymptote to anti-de~Sitter (AdS) spacetime, also asymptotic flatness does not hold anymore.
In this case, the horizon of a black hole can indeed be a compact Riemann surface $\Sigma_g$ of any
genus $g$ \cite{Lemos:1994xp,Mann:1996gj,Vanzo:1997gw,Cai:1996eg}.
It should be noted that, unless $g=0$, these spacetimes are asymptotically only locally
AdS; their global structure is different. This is in contrast to the black rings in five dimensions,
which are asymptotically Minkowski, in spite of their nontrivial horizon topology.
Notice, in addition, that the solutions of \cite{Lemos:1994xp,Mann:1996gj,Vanzo:1997gw,Cai:1996eg}
do not exhaust the spectrum of black holes in AdS$_4$,
since one can also have horizons that are noncompact manifolds with yet finite area (and thus
finite entropy), topologically spheres with two punctures
\cite{Gnecchi:2013mja,Klemm:2014rda}\footnote{These solutions can be generalized to
$D>4$ \cite{Hennigar:2014cfa}.}.

In this paper, we will allow for both of the possibilities described above, i.e., we shall consider the
case $D=5$ and include a negative cosmological constant. More generally, our model contains
scalar fields with a potential that admits AdS$_5$ vacua. A class of uncharged black holes in
Einstein-Lambda gravity was obtained by Birmingham in \cite{Birmingham:1998nr} for arbitrary
dimension $D$. These solutions have the property that the horizon is a $(D-2)$-dimensional Einstein manifold of positive, zero or negative curvature. In our case, $D=5$, and three-dimensional
Einstein spaces have necessarily constant curvature, i.e., are homogeneous and isotropic.
Similar to what is done in Bianchi cosmology, one can try to relax these conditions by dropping
the isotropy assumption. The horizon is then a homogeneous manifold, and belongs thus to the nine
``Bianchi cosmologies'', which are in correspondence with the eight Thurston model geometries,
cf.~appendix \ref{app:homogeneous-spaces} for details. For two of these cases, namely Nil and Sol,
the corresponding black holes in five-dimensional gravity with negative cosmological constant
were constructed in \cite{Cadeau:2000tj} for the first time. Asymptotically, these solutions are neither
flat nor AdS, but exhibit anisotropic scaling. Ref.~\cite{Faedo:2019rgo} went one step further with
respect to \cite{Cadeau:2000tj} by adding also charge and studying the attractor mechanism for black
holes with homogeneous but anisotropic horizons.
Additional charged backgrounds were constructed, e.g., in~\cite{Arias:2017yqj}, where an intrinsically dyonic black hole with Sol horizon in Einstein-Maxwell-AdS gravity (with no Chern-Simons term) was found, and in~\cite{Bravo-Gaete:2017nkp}, which considers different models that are not directly related to gauged supergravity theories.
Moreover, higher-dimensional (uncharged) generalizations were obtained in \cite{Hervik:2003vx,Hervik:2007zz}.

Up to now, neither supersymmetric nor rotating solutions of this type have been known, with the only
exception being the one obtained in \cite{Gutowski:2004ez} by a certain scaling limit from a spherical
black hole. For the homogeneous space $\SL(2,\RR$), the corresponding black hole is not known even in
the static uncharged case, and to partially fill these gaps is the scope of the present paper.
There are various motivations for this. First of all, one may wish to further explore the full solution space
of the Einstein-Maxwell equations with negative cosmological constant in five dimensions. This is
interesting since such configurations can potentially be lifted to ten dimensions and represent thus
possible superstring vacua, but also from a holographic point of view, especially in condensed matter
applications of holography, due to the anisotropic scaling of these solutions.
Moreover, black holes with horizons other than spherical are intriguing in their own right, since
in such contexts new features of black hole physics may emerge that have been hitherto unknown.

We start in section \ref{sec:gauged-sugra} by setting up the minimal gauged supergravity model that
will be considered throughout the major part of this paper, and briefly review the recipe of
\cite{Gauntlett:2003fk} that allows for a systematic construction of timelike supersymmetric backgrounds,
starting from a given K\"ahler base space. In subsection \ref{subsec:BPS-Nil} we use this to directly
rederive the Nil solution of \cite{Gutowski:2004ez} (obtained from a spherical black hole in a limit
when the latter is very large), and to show what the corresponding K\"ahler base is.
At the level of the isometry groups, the scaling limit of \cite{Gutowski:2004ez}
is easy to understand: Nil can be represented as the Lie group of
$3\times 3$ upper triangular matrices of the form
\eq
M = \left(\begin{array}{ccc} 1 & x & z \\ 0 & 1 & y \\ 0 & 0 & 1\end{array}\right)\,,
\end{equation}
with $x,y,z\in\mathbb{R}$. The Lie algebra generators are thus
\eq
L_x = \left(\begin{array}{ccc} 0 & 1 & 0 \\ 0 & 0 & 0 \\ 0 & 0 & 0\end{array}\right)\,, \qquad
L_y = \left(\begin{array}{ccc} 0 & 0 & 0 \\ 0 & 0 & 1 \\ 0 & 0 & 0\end{array}\right)\,, \qquad
L_z = \left(\begin{array}{ccc} 0 & 0 & 1 \\ 0 & 0 & 0 \\ 0 & 0 & 0\end{array}\right)\,,
\feq
which obey
\eq
[L_x, L_y] = L_z\,, \qquad [L_x, L_z] = [L_y, L_z] = 0\,, \label{Lie-alg-Nil}
\feq
which is of course the Heisenberg algebra with central element $L_z$. On the other hand,
the Berger- (or squashed) sphere\footnote{Below it will become clear why one cannot
start from the round S$^3$.} has isometry group $\text{SU}(2)\times\text{U}(1)$, with the $\text{SU}(2)$
generators satisfying
\eq
[J_x, J_y] = J_z\,, \qquad [J_y, J_z] = J_x\,, \qquad [J_z, J_x] = J_y\,. \label{Lie-alg-SU(2)}
\feq
Now rescale $J_{x,y}\to\lambda J_{x,y}$, $J_z\to\lambda^2 J_z$, and take the limit $\lambda\to\infty$.
This is an In\"on\"u-Wigner contraction that transforms \eqref{Lie-alg-SU(2)} into \eqref{Lie-alg-Nil}.\\
The metric on the Berger sphere can be written as
\eq
ds^2 = a^2 (d\theta^2 + \sin^2\theta d\psi^2) + b^2 (d\phi + \cos\theta d\psi)^2\,, \label{metr-Berger}
\feq
with the two radii $a,b$. Now set
\eq
x = \lambda\theta\cos\psi\,, \qquad y = \lambda\theta\sin\psi\,, \qquad z = \lambda^2 (\phi + \psi)\,,
\feq
and take the limit $\lambda\to\infty$ while keeping $x,y,z$ fixed. This means $\theta\to 0$, so that
we effectively zoom onto the north pole of the $\text{S}^2$ over which the Berger sphere is fibered.
Using $\cos\theta=1-2\sin^2(\theta/2)$, it is easy to see that \eqref{metr-Berger} has a finite limit
for $\lambda\to\infty$ if we also set
\eq
A=a/\lambda\,, \qquad B=b/\lambda^2\,, \label{scaling-radii}
\feq
and keep $A,B$ fixed. From \eqref{scaling-radii} it is evident that $a$ and $b$ must scale differently,
and thus we cannot start from the round $\text{S}^3$ since the latter has $a=b$. Then, the limit
of \eqref{metr-Berger} for $\lambda\to\infty$ is given by
\begin{equation} \label{metr-Nil}
ds^2 = A^2 (dx^2 + dy^2) + B^2 \left(dz + \frac{y dx - x dy}{2}\right)^2\,,
\end{equation}
which is the metric on the homogeneous space Nil, invariant under the transformations
\eq
z\mapsto z + a\,; \qquad x\mapsto x + b\,, \quad z\mapsto z + \frac b2 y\,; \qquad y\mapsto y + c\,,
\quad z\mapsto z - \frac c2 x\,,
\feq
that generate the Heisenberg algebra. An additional isometry consists of the $\text{U}(1)$ rotations
in the $(x,y)$-plane that leave of course the area form $dx\wedge dy=d[\frac12(xdy-ydx)]$
invariant.

In subsection \ref{subsec:Nonextr-Nil} we apply a scaling limit similar in spirit to that of
\cite{Gutowski:2004ez} to the Kerr-AdS$_5$ solution with two equal rotation parameters, which consists
again in zooming onto the north pole of the S$^2$ over which the S$^3$ is fibered, while boosting the
horizon velocity effectively to the speed of light. This leads to a nonextremal rotating Nil black hole
that was not known before.

In section \ref{sec:SL} we use again the techniques of \cite{Gauntlett:2003fk} to construct the first
example of a black hole with $\SL(2,\RR$) horizon. It turns out to be rotating and one quarter BPS. The
physical properties of this solution are discussed, and it is shown that in the near-horizon limit it boils
down to the geometry of \cite{Gutowski:2004ez}, with a supersymmetry enhancement to one half.
Dimensional reduction to $D=4$ gives a new solution with hyperbolic horizon to the t$^3$ model that
carries both electric and magnetic charges.
Finally, in \ref{sec:matter-coupled}, the black hole of section \ref{sec:SL} is generalized to
$N=2$, $D=5$ gauged supergravity coupled to an arbitrary number of abelian vector multiplets,
and subsequently illustrated for the stu model.
We conclude in \ref{sec:final-rem} with some final remarks. Appendix \ref{app:homogeneous-spaces}
contains a summary of some material on homogeneous manifolds.

\section{$N=2$, $D=5$ minimal gauged supergravity}
\label{sec:gauged-sugra}

Let us consider $N=2$, $D=5$ minimal gauged supergravity, whose bosonic field content is given by the supergravity multiplet only, which includes the f\"unfbein $e^a_\mu$ and the graviphoton $A_\mu$. In the case of minimal theory, i.e.\ in absence of matter multiplets, the abelian gauging contributes to the action with a cosmological constant $\Lambda=-6/\ell^2$ and a Chern-Simons term for the graviphoton. The action reads
\begin{equation} \label{action}
	\mathcal{S} = \frac{1}{16\pi G} \int \, \biggl[ (R - 2 \Lambda) \star\!1 - 2 \star\!F \wedge F - \frac{8}{3\sqrt{3}} F \wedge F \wedge A \biggr] \,,
\end{equation}
where $F= dA$ is the field strength of the U(1) gauge field. The equations of motion follow immediately
\begin{equation}
	\begin{gathered}
		R_{\mu\nu} - 2F_{\mu\lambda} F_\nu^{\ \lambda} + \frac13 \, g_{\mu\nu} \biggl( F_{\rho\sigma} F^{\rho\sigma} + \frac{12}{\ell^2} \biggr) = 0 \,, \\
		d \star\!F + \frac{2}{\sqrt{3}} F \wedge F = 0 \,.
	\end{gathered}
\end{equation}
Any supersymmetric solution of the equations of motion must admit a Killing spinor, from which it is possible to construct a real scalar $f$, a real vector $V$ and three real two-forms $J^{(i)}$, $i=1,2,3$, that satisfy the following algebraic relations~\cite{Gauntlett:2002nw,Gutowski:2004ez}:
\begin{subequations}
\begin{gather}
V^\mu V_\mu = -f^2 \,, \label{V2-f2} \\
J^{(i)} \wedge J^{(j)} = -2 \delta_{ij} f \star\!V \,, \\
i_V J^{(i)} = 0 \,, \label{hyperK1} \\
i_V \star\!J^{(i)} = -f J^{(i)} \,, \\
(J^{(i)})_\mu^{\,\ \rho} (J^{(j)})_\rho^{\,\ \nu} = -\delta_{ij} \bigl(f^2 \delta^\nu_\mu + V_\mu V^\nu\bigr) + \varepsilon_{ijk} f(J^{(k)})_\mu^{\,\ \nu} \,, \label{hyperK3}
\end{gather}
\end{subequations}
where $\varepsilon_{123}=+1$ and $i_V$ denotes the interior product with the vector $V$.
From~\eqref{V2-f2} we see that $V$ can be either timelike or null. $f$, $V$ and $J^{(i)}$ must also satisfy the differential relations~\cite{Gauntlett:2003fk,Gutowski:2004ez}
\begin{subequations}
\begin{gather}
df = -\frac{2}{\sqrt{3}} i_V F \,, \label{eq:df} \\
\nabla_{(\mu} V_{\nu)} = 0 \,, \\[0.25em]
dV = -\frac{4}{\sqrt{3}} f F - \frac{2}{\sqrt{3}} \star\!(F \wedge V) - \frac{2}{\ell} J^{(1)} \,,  \label{eq:dV} \\
dJ^{(i)} = \frac{1}{\ell} \, \varepsilon_{1ij} \bigl[ 2\sqrt{3} A \wedge J^{(j)} + 3 \star\!J^{(j)} \bigr] \,.
\end{gather}
\end{subequations}
The first two relations imply that $V$ is a Killing vector that leaves the field strength invariant, thus generating a symmetry of the solution. From the last one we see that $J^{(1)}$ is closed.

In what follows we shall focus on the case in which $V$ is not globally null, and thus it is possible to find an open set $\mathcal{U}$ in which $V$ is timelike. Assuming, without loss of generality, $f>0$ in $\mathcal{U}$, we can introduce a set of coordinates $(t,x^m)$ such that we can make an ansatz on the metric of the form
\begin{equation} \label{timelike-metric}
	ds^2 = -f^2 (dt + \omega)^2 + f^{-1} ds_4^2 \,,
\end{equation}
where $V=\partial_t$, $ds_4^2$ is the line element of a four-dimensional Riemannian ``base space''~$\mathcal{B}$ orthogonal to the orbits of $V$ and $\omega$ is a one-form defined on $\mathcal{B}$. Relations \eqref{hyperK1}--\eqref{hyperK3}, together with supersymmetry, imply that $\mathcal{B}$ is K\"ahler, with anti-selfdual K\"ahler form~$J^{(1)}$, once $J^{(1)}$ is restricted to~$\mathcal{B}$ \cite{Gauntlett:2003fk}.

The strategy of~\cite{Gutowski:2004ez}, to which we refer for a deeper analysis (see also~\cite{Gauntlett:2002nw,Gauntlett:2003fk}), consists in reducing necessary and sufficient conditions for supersymmetry to a set of equations that will allow us to determine all the functions appearing in the metric.
Following~\cite{Gutowski:2004ez}, we begin with the field strength $F$. Equations~\eqref{eq:df} and~\eqref{eq:dV} imply
\begin{equation}
	F = \frac{\sqrt3}{2} \, d[f(dt + \omega)] - \frac{1}{\sqrt3} G^+ - \frac{\sqrt3}{\ell} f^{-1} J^{(1)} \,,
\end{equation}
where we defined ($\star_4$ denotes the Hodge dual on $\mathcal{B}$)
\begin{equation} \label{eq:G+-}
	G^\pm \equiv \frac{f}{2} \, (d\omega \pm \star_4 d\omega) \,.
\end{equation}
Function $f$ is fixed by supersymmetry to
\begin{equation} \label{eq:f}
	f^{-1} = -\frac{\ell^2 R}{24} \,,
\end{equation}
where $R$ is the Ricci scalar of $\mathcal{B}$.
Supersymmetry conditons and Maxwell equations yield the following equations (latin indices denote curved indices on~$\mathcal{B}$):
\begin{gather}
	\label{eq:G+}
	G^+ = -\frac{\ell}{2} \Bigl(\mathcal{R} - \frac{R}{4} J^{(1)}\Bigr) \,, \\
	\label{eq:PDEf}
	\nabla^2 f^{-1} = \frac29 (G^+)^{mn} (G^+)_{mn} + \frac{1}{\ell} f^{-1} (G^-)^{mn} (J^{(1)})_{mn} - \frac{8}{\ell^2} f^{-2} \,.
\end{gather}
Here, $\mathcal{R}$ is the Ricci form on $\mathcal{B}$ defined as
\begin{equation} \label{ricci-form}
	\mathcal{R}_{mn} \equiv \frac{1}{2} R_{mnpq} (J^{(1)})^{pq} \,,
\end{equation}
and $R_{mnpq}$ and $\nabla^2$ are, respectively, the Riemann tensor and the Laplacian on $\mathcal{B}$.
Definitions~\eqref{eq:G+-} imply that the two-form $f^{-1}(G^+ + G^-)$ must be closed.
As it was pointed out in~\cite{Figueras:2006xx}, this requirement gives a nontrivial constraint on the K\"ahler base space. The authors of~\cite{Cassani:2015upa} further elaborated on this condition, obtaining the following constraint that $\mathcal{B}$ must satisfy:
\begin{equation} \label{kahler-constr}
	\nabla^2 \biggl( \frac12 \nabla^2 R + \frac23 R^{mn} R_{mn} - \frac13 R^2 \biggr) + \nabla_m (R^{mn} \partial_n R) = 0\,.
\end{equation}
Lastly, by means of~\eqref{eq:f} and~\eqref{eq:G+}, the field strength takes the form
\begin{equation} \label{field-strength}
	F = \frac{\sqrt{3}}{2} \, d[f (dt + \omega)] + \frac{\ell}{2\sqrt{3}} \mathcal{R} \,.
\end{equation}
Summarizing, the main steps to construct new timelike supersymmetric solutions to five-dimensional minimal gauged supergravity are the following.
Start choosing a K\"ahler space $\mathcal{B}$ with negative curvature, K\"ahler two-form $J^{(1)}$ and satisfying the constraint~\eqref{kahler-constr}. Determine $f$ and~$\omega$ by means of equations \eqref{eq:f}, \eqref{eq:G+} and \eqref{eq:PDEf}, thus obtaining the five-dimensional metric~\eqref{timelike-metric}. The field strength is given by~\eqref{field-strength}.
Remarkably, a solution to these equations always exists~\cite{Gauntlett:2003fk} and the resulting system preserves at least one quarter of the total number of supersymmetries~\cite{Gutowski:2004ez}.

\section{Nil black holes}
\label{sec:Nil}

\subsection{BPS case}
\label{subsec:BPS-Nil}

As we explained in the introduction, the authors of \cite{Gutowski:2004ez} applied a certain scaling
limit to a black hole with spherical horizon to obtain a solution with spatial cross-sections modelled on
nilgeometry. In this section we shall directly rederive the latter by applying the recipe of
\cite{Gutowski:2004ez}. This computation will serve as a warm-up for the construction of black holes with $\SL(2,\RR)$ horizon, which will be dealt with in the next section.\\
The starting point of our construction is the K\"ahler base space of the near-horizon
limit \cite{Gutowski:2004ez},
\begin{equation} \label{Nil_base-nh}
	ds^2_4 = d\rho^2 + \frac{12\rho^2}{\ell^2}\biggl[dz' + \frac{\sqrt3}{2\ell}(y dx - x dy)\biggr]^2 + \frac{3\rho^2}{\ell^2}\bigl(dx^2 + dy^2\bigr)\,.
\end{equation}
Inspired by \cite{Gutowski:2004ez} and by the metric \eqref{Nil_base-nh}, we make the following ansatz for the base space and the corresponding K\"ahler two-form $J^{(1)}$ in the case of full black hole:
\begin{equation}
	\begin{aligned}
		ds_4^2 &= d\rho^2 + a(\rho)^2\bigl[(\sigma_L^1)^2 + (\sigma_L^2)^2\bigr] + b(\rho)^2 (\sigma_L^3)^2\,, \\
		J^{(1)} &= d[c(\rho)\sigma_L^3]\,,
	\end{aligned}
\end{equation}
where the $\sigma_L^i$ are the Nil-invariant one-forms \eqref{nil_forms}. Assuming $a,b>0$ we introduce the orthonormal basis
\begin{equation}
	e^1 = d\rho\,, \qquad e^2 = a\sigma_L^1\,, \qquad e^3 = a\sigma_L^2\,, \qquad e^4 = b\sigma_L^3\,.
\end{equation}
Requiring $J^{(1)}$ to be an anti-selfdual complex structure implies
\begin{equation} \label{Nil_abc}
	c = -\epsilon a^2\,, \qquad b = 2aa'\,,
\end{equation}
where $\epsilon=\pm1$. Thus, we have
\begin{equation}
	J^{(1)} = -\epsilon (e^1\wedge e^4 - e^2\wedge e^3)\,.
\end{equation}
From equ.~\eqref{eq:f} we can write $f$ in terms of $a$ as
\begin{equation}
f^{-1} = \frac{\ell^2}{12 a^2 a'}\bigl(4(a')^3 + 7a a' a'' + a^2 a'''\bigr)\,.
\end{equation}
For the one-form $\omega$ we choose an ansatz analogous to \cite{Gutowski:2004ez},
\begin{equation}
\omega = \Psi(\rho)\sigma_L^3\,.
\end{equation}
Plugging this into \eqref{eq:G+-} yields\footnote{In our construction, in which $G^\pm$ are defined starting from an ansatz on $\omega$, constraint~\eqref{kahler-constr} is automatically satisfied once all the other equations are solved.}
\begin{equation}
G^{\pm} = \frac f2\biggl(\frac{\Psi'}{2a a'}\mp\frac{\Psi}{a^2}\biggr) (e^1 \wedge e^4 \pm e^2 \wedge e^3)\,.
\end{equation}
Comparing this with \eqref{eq:G+}, one gets
\begin{equation}\label{eq:psi'-psi}
\frac{\Psi'}{2a a'} - \frac{\Psi}{a^2} = \frac{\epsilon\ell g}{2f}\,,
\end{equation}
where, for brevity,
\begin{equation}
g = \frac{4(a')^3 - 3a a' a'' - a^2 a'''}{a^2 a'}\,.
\end{equation}
Inserting the expression for $G^{\pm}$ into \eqref{eq:PDEf} we obtain
\begin{equation}
\label{eq:psi'-psi-f}
\frac{\Psi'}{2a a'} + \frac{\Psi}{a^2} = -\frac{\epsilon\ell}2\Bigl(\nabla^2 f^{-1} + \frac8{\ell^2} f^{-2} -
\frac{\ell^2 g^2}{18}\Bigr)\,.
\end{equation}
For our ansatz, it is straightforward to shew that
\begin{equation}
\nabla^2 f^{-1} = \frac1{a^3 a'}\partial_{\rho}\bigl(a^3 a'\partial_{\rho} f^{-1}\bigr)\,.
\end{equation}
Taking the difference between \eqref{eq:psi'-psi-f} and \eqref{eq:psi'-psi} allows to eliminate $\Psi'$,
\begin{equation}\label{Nil_Psi}
\Psi = -\frac{\epsilon\ell a^2}4\Bigl(\nabla^2 f^{-1} + \frac8{\ell^2} f^{-2} - \frac{\ell^2 g^2}{18} +
f^{-1} g\Bigr)\,.
\end{equation}
Using this in \eqref{eq:psi'-psi} gives the master equation
\begin{equation} \label{eq:PDEa}
\Bigl(\nabla^2 f^{-1} + \frac8{\ell^2} f^{-2} - \frac{\ell^2 g^2}{18} + f^{-1} g\Bigr)' + \frac{4a'}a f^{-1} g
= 0\,.
\end{equation}
This is a sixth order nonlinear ordinary differential equation for $a(\rho)$. Solving it will allow us to write down both the black hole metric and the Maxwell field strength.
Following \cite{Gutowski:2004ez}, we make an ansatz for the function $a$ of the form $a=\alpha\sinh(\beta\rho/\ell)$. This solves \eqref{eq:PDEa} for generic values of $\alpha$ and $\beta$. By rescaling the coordinates one can set $\beta=1$ without loss of generality. Moreover, matching the $\rho\to0$ limit of our solution and the near-horizon base space \eqref{Nil_base-nh} fixes $\alpha=\sqrt3$. At the end, we obtain
\begin{equation}
a = \sqrt3 \sinh(\rho/\ell)\,,
\end{equation}
and the other functions read
\begin{equation}
b = \frac3{\ell} \sinh(2\rho/\ell)\,, \quad  f = \frac{3\sinh^2(\rho/\ell)}{3\sinh^2(\rho/\ell) + 1}\,,
\quad\Psi = -\epsilon\frac{3\sinh^2(2\rho/\ell) + 2}{2\ell\sinh^2(\rho/\ell)}\,.
\end{equation}
The full black hole solution can be written in a neater fashion by defining the new radial coordinate
\begin{equation}
	R = \sqrt{3\sinh^2(\rho/\ell) + 1}\,,
\end{equation}
which gives
\begin{equation}
f(R) = 1 - R^{-2}\,, \qquad\Psi(R) = -\frac{2\epsilon R^2}{\ell} f(R)^{-1}\left(1 + \frac1{R^2} -
\frac1{2R^4}\right)\,.
\end{equation}
The metric and Maxwell gauge potential are then
\begin{align}
	ds^2 &= -f^2 dt^2 - 2 f^2\Psi\sigma_L^3 dt + \frac{\ell^2 dR^2}{f^2 (R^2 + 2)} + R^2
	\left[(\sigma_L^1)^2 + (\sigma_L^2)^2 + \frac{4R^2 - 1}{\ell^2 R^6}(\sigma_L^3)^2\right]\,, \nonumber \\
	A &= \frac{\sqrt3}2\left(f dt + \frac{\epsilon}{\ell R^2}\sigma_L^3\right)\,, \label{metr-A-Nil}
\end{align}
which is precisely the solution obtained in section 4.4 of \cite{Gutowski:2004ez} as a scaling limit of a
spherical black hole, with the radial coordinate called $S$ and $z$ rescaled by $\sqrt3/\ell$.

\subsection{Nonextremal black holes with Nil horizon from Kerr-AdS$_5$}
\label{subsec:Nonextr-Nil}

In this subsection we shall apply a scaling limit similar in spirit to that of \cite{Gutowski:2004ez}
to the Kerr-AdS$_5$ solution, that amounts to a group contraction from S$^3$ to Nil.
We start from KAdS$_5$ with two equal rotation parameters, which is given by \cite{Hawking:1998kw}
\begin{align}
ds^2 &= -\left[1 + \frac{r^2 + a^2}{\ell^2} -\frac{2M}{r^2 + a^2}\right] dt^2 + \frac{r^2 + a^2}{\Delta}
dr^2 + \frac a\Xi\left[\frac{r^2 + a^2}{\ell^2} - \frac{2M}{r^2 + a^2}\right]\sigma_L^3 dt\nonumber \\
& \phantom{{}=} + \frac{r^2 + a^2}{4\Xi}\left((\sigma_L^1)^2 + (\sigma_L^2)^2\right) + \left[\frac{r^2 + a^2}{4\Xi}
+ \frac{M a^2}{2\Xi^2 (r^2 + a^2)}\right](\sigma_L^3)^2\,, \label{KAdS5}
\end{align}
where
\begin{equation}
\Delta = \frac1{r^2}(r^2 + a^2)^2\left(1 + \frac{r^2}{\ell^2}\right) - 2M\,, \qquad \Xi = 1 -
\frac{a^2}{\ell^2}\,,
\end{equation}
and $M$, $a$ denote the mass and rotation parameters respectively. To have the correct signature,
one needs $\Xi>0$, and thus $|a|<\ell$.
For convenience, we have rewritten the solution of \cite{Hawking:1998kw} in terms of right-invariant
one-forms\footnote{The angles in \eqref{right-inv-S^3} are related to those of \cite{Hawking:1998kw}
by $\theta=2\theta_{\text{there}}$, $\phi=\psi_{\text{there}}+\phi_{\text{there}}$,
$\psi=\psi_{\text{there}}-\phi_{\text{there}}$.}
\begin{equation}
\begin{split}
\sigma_L^1 &= \sin\phi d\theta - \cos\phi\sin\theta d\psi\,, \\
\sigma_L^2 &= \cos\phi d\theta + \sin\phi\sin\theta d\psi\,, \label{right-inv-S^3} \\
\sigma_L^3 &= d\phi + \cos\theta d\psi\,,
\end{split}
\end{equation}
on S$^3$.

It turns out that, to obtain a Nil black hole from \eqref{KAdS5}, one has to combine the limit of
\cite{Gutowski:2004ez} with the one taken in \cite{Caldarelli:2008pz,Caldarelli:2011idw}, where
the horizon rotates effectively at the speed of light, $a\to\ell$.
Specifically, we introduce the new coordinates
\begin{equation}
x = \Xi^{-1/2}\theta\cos\psi\,, \qquad y = \Xi^{-1/2}\theta\sin\psi\,, \qquad z = \Xi^{-1}(\phi + \psi)\,,
\end{equation}
and then scale $a\to\ell$, i.e., $\Xi\to 0$, keeping $x$, $y$ and $z$ fixed. This means that the angle
$\theta$ must go to zero, which implies that we effectively zoom onto the north pole of the S$^2$, over
which the S$^3$ is fibered. Defining moreover $\rho^2\equiv r^2+\ell^2$, the resulting new metric is
given by
\begin{equation} \label{metr-Nil-nonextr-rot}
\begin{split}
ds^2 &= -\left[1 + \frac{\rho^2}{\ell^2} - \frac{2M}{\rho^2}\right] dt^2 + \frac{d\rho^2}{\frac{\rho^2}
{\ell^2} - \frac{2M}{\rho^2} + \frac{2M\ell^2}{\rho^4}} + \frac{\rho^2}4 (dx^2 + dy^2) \\
& \phantom{{}=} + \left[\frac{\rho^2}{\ell} - \frac{2M\ell}{\rho^2}\right] \left(dz + \frac{y dx - x dy}{2}\right) dt
+ \frac{M\ell^2}{2\rho^2}\left(dz + \frac{y dx - x dy}{2}\right)^2\,.
\end{split}
\end{equation}
We have checked that \eqref{metr-Nil-nonextr-rot} satisfies the Einstein equations with negative
cosmological constant $\Lambda=-6/\ell^2$.

\eqref{metr-Nil-nonextr-rot} has a Nil geometry horizon at the largest root of
\begin{equation}
\frac{\rho^2}{\ell^2} - \frac{2M}{\rho^2} + \frac{2M\ell^2}{\rho^4} = 0\,.
\end{equation}
For $M<27\ell^2/8$ there is a naked singularity, in the case $M=27\ell^2/8$ we have an extremal
black hole, while for $M>27\ell^2/8$ the Hawking temperature is nonvanishing.
In addition, there is an ergosphere at $g_{tt}=0$, i.e.,
\begin{equation}
\frac{\rho_{\text{erg}}^2}{\ell^2} = \frac12\left(\sqrt{1 + \frac{8M}{\ell^2}} - 1\right)\,.
\end{equation}
In the extremal case, the horizon is at $\rho^2_{\text{hor}}/\ell^2=3/2$ and the ergosphere at
$\rho^2_{\text{erg}}/\ell^2=\sqrt7-1/2$. Presumably, the solution \eqref{metr-Nil-nonextr-rot} is a
special case of a more general black hole (still to be constructed) with independent values of mass and angular momentum~$J$, where $J$ has been fixed to a certain value. If so, the condition $M>27\ell^2/8$, necessary to avoid naked singularities, results probably from a more general bound that involves both $M$ and $J$.

Notice that the radial dependence of \eqref{metr-Nil-nonextr-rot} is qualitatively very different from that of the static Cadeau-Woolgar solution (cf.~equ.~(II.23) of \cite{Cadeau:2000tj}), which makes it hard to guess an ansatz that contains both (II.23) of \cite{Cadeau:2000tj} and \eqref{metr-Nil-nonextr-rot}.

It would be interesting to apply a similar scaling limit to KAdS$_5$ with two unequal
rotation parameters and to its charged extension constructed in~\cite{Chong:2005hr}.
We shall postpone this to a future publication.

\section{SL(2,\,$\RR$) black hole}
\label{sec:SL}

The aim of this section is to construct a supersymmetric black hole whose horizon is modelled on the
$\SL(2,\RR)$ geometry, using the recipe originally proposed in \cite{Gauntlett:2003fk}.
The ans\"atze for the base space and the K\"ahler form $J^{(1)}$ are the same as before,
\begin{equation}
	\begin{aligned}
		ds_4^2 &= d\rho^2 + a(\rho)^2\bigl[(\sigma_L^1)^2 + (\sigma_L^2)^2\bigr] + b(\rho)^2 (\sigma_L^3)^2\,, \\
		J^{(1)} &= d[c(\rho)\sigma_L^3]\,,
	\end{aligned}
\end{equation}
with the $\SL(2,\RR)$-invariant base one-forms given in \eqref{sl_forms}.
Following \cite{Gauntlett:2003fk}, where the same was done for a family of black holes with spherical horizon (cf.~also the review in section \ref{sec:Nil} for the Nil-invariant case), we find that the whole system is completely described by the sixth-order differential equation for $a(\rho)$
\begin{equation} \label{SL_master-eq}
\Bigl(\nabla^2 f^{-1} + \frac8{\ell^2} f^{-2} - \frac{\ell^2 g^2}{18} + f^{-1} g\Bigr)' + \frac{4a'}a f^{-1} g
= 0\,,
\end{equation}
with
\begin{equation}
	\begin{aligned}
		f^{-1} &= \frac{\ell^2}{12 a^2 a'}\bigl(4(a')^3 + 7a a' a'' + a^2 a''' + a' \bigr)\,, \\
		g &= \frac{4(a')^3 - 3a a' a'' - a^2 a''' + a'}{a^2 a'}\,.
	\end{aligned}
\end{equation}
As in \cite{Gutowski:2004ez}, \eqref{SL_master-eq} is again solved by the ansatz
\begin{equation} \label{eq:a-SL}
	a = \alpha\ell\sinh(\rho/\ell)\,,
\end{equation}
where the value of $\alpha$ is fixed to
\begin{equation}
	\alpha = \frac12\sqrt{\frac{9\ell^{-2} + \Delta^2}{3\ell^{-2} - \Delta^2}}
\end{equation}
in order to have the correct matching with the near-horizon base space (3.65) of \cite{Gutowski:2004ez}.
Since $0<\Delta<\sqrt3/\ell$ \cite{Gutowski:2004ez}, one has $\alpha>\sqrt3/2$. The expressions for
$b$ and $c$ follow directly from \eqref{Nil_abc}, while $\Psi$ can be read from \eqref{Nil_Psi}.

For what follows, it will be useful to introduce a new radial coordinate $R$ as
\begin{equation}
	R^2 = 4 f^{-1} a^2 = \ell^2\left[4\alpha^2\sinh^2(\rho/\ell) + \frac{4\alpha^2 + 1}3 \right]\,.
\end{equation}
It is convenient to define the constant
\begin{equation}
	R_0 = \ell\sqrt{\frac{4\alpha^2 + 1}3}\,,
\end{equation}
which satisfies
\begin{equation} \label{constr-R_0}
	R_0 > \frac{2\ell}{\sqrt3}
\end{equation}
due to $\alpha>\sqrt3/2$. We then have
\begin{equation} \label{SL_f+Psi}
f(R) = 1 - \frac{R_0^2}{R^2}\,, \qquad\Psi(R) = -\frac{\epsilon R^2}{2\ell} f(R)^{-1}\left(1 +
\frac{R_0^2}{R^2} - \frac{R_0^4}{2R^4} \right)\,,
\end{equation}
and the black hole metric can be written as
\begin{equation} \label{SL_metric}
ds^2 = -f^2 dt^2 - 2 f^2\Psi\sigma_L^3 dt + \frac{dR^2}U + \frac{R^2}4\bigl[(\sigma_L^1)^2 + (\sigma_L^2)^2 + V (\sigma_L^3)^2\bigr]\,,
\end{equation}
where
\begin{equation} \label{SL_U+V}
U(R) = f(R)^2\left(-1 + \frac{2R_0^2}{\ell^2} + \frac{R^2}{\ell^2}\right)\,, \qquad V(R) = -1 +
\frac{R_0^6}{\ell^2 R^4} - \frac{R_0^8}{4\ell^2R^6}\,.
\end{equation}
In these coordinates the gauge potential becomes
\begin{equation} \label{SL_gauge}
A = \frac{\sqrt3}2\biggl(f dt + \frac{\epsilon R_0^4}{4\ell R^2}\sigma_L^3\biggr)\,.
\end{equation}
Similar to what happens for the spherical solution constructed in \cite{Gutowski:2004ez}, in the large black hole limit $R_0\to\infty$, \eqref{SL_metric} and \eqref{SL_gauge} should again boil down to the configuration \eqref{metr-A-Nil} with Nil horizon, although we did not check this explicitely.

\subsection{Physical discussion}

In this subsection we shall analyse the physics of the $\SL(2,\RR)$-invariant family of solutions \eqref{SL_metric}.
There is a curvature singularity in $R=0$, where the Kretschmann scalar diverges, and a regular horizon for $f(R)=0$, i.e., $R=R_0$. This coordinate singularity can be eliminated by introducing Gaussian null coordinates, that allow to extend the metric beyond the horizon. Defining $u$, $z'$ and $r$ by
\begin{equation}
dt = du - \frac{V}{U} dr\,, \qquad dz = dz' - \frac{4f^2\Psi}{R^2 U} dr\,, \qquad dR = V^{1/2} dr\,,
\end{equation}
with $R=R_0$ at $r=0$, \eqref{SL_metric} becomes
\begin{equation}
	ds^2 = -f^2 du^2 + 2du dr - 2f^2\Psi {\sigma_L^3}' du + \frac{R^2}4\bigl[(\sigma_L^1)^2 + (\sigma_L^2)^2 + V ({\sigma_L^3}')^2\bigr]\,,
\end{equation}
where ${\sigma_L^3}'$ is defined as $\sigma_L^3$ in \eqref{sl_forms}, but with $z'$ in place of $z$.
One easily shows that $f\sim r$ for small values of $r$, thus both $f^2$ and $f^2\Psi$ smoothly go to zero when $r\to0$, leaving a nonsingular metric on the horizon. Note that the gauge potential \eqref{SL_gauge} becomes singular in Gaussian coordinates, but this apparent pathology can be eliminated by means of a suitable gauge transformation $A\mapsto A+d\lambda$, with $\lambda=\lambda(r)$.

As expected for a supersymmetric (and thus extremal) black hole, its near-horizon geometry contains an AdS$_2$ factor, mixed with the $\SL(2,\RR)$-invariant slices due to the rotation. To see this, introduce
the new coordinates $(T,\zeta)$ according to
\begin{equation}
	t = \frac{T}{\varepsilon}\,, \qquad R = R_0 (1 + \varepsilon\zeta)
\end{equation}
in \eqref{SL_metric}, and take the limit $\varepsilon\to 0$, which leads to the near-horizon metric
\begin{equation}\label{SL_metric-nh}
ds^2 = -\frac{4W}{V_0}\zeta^2 dT^2 + \frac{R_0^2 d\zeta^2}{4W\zeta^2} + \frac{R_0^2}4
\biggl[(\sigma_L^1)^2 + (\sigma_L^2)^2 + V_0\left(\sigma_L^3 + \frac{6\epsilon}{\ell V_0}\zeta dT\right)^2
\,\biggr]\,,
\end{equation}
where we defined
\begin{equation}
	V_0 = V(R_0) = -1 + \frac{3R_0^2}{4\ell^2}\,, \qquad W = -1 + \frac{3R_0^2}{\ell^2}\,.
\end{equation}
Notice that the correct signature is ensured by the constraint \eqref{constr-R_0}, which implies $W>0$, $V_0>0$.

The computation of the physical quantities like mass and angular momenta is a nontrivial issue,
since the standard Komar integrals associated to the Killing vectors $\partial_t$, $\partial_y$ and
$\partial_z$ diverge due to the presence of a vacuum energy related to the cosmological constant.
Moreover, there is no obvious background to be subtracted, and other techniques, like the
Ashtekar-Magnon-Das formalism \cite{Ashtekar:1984zz,Ashtekar:1999jx}, cannot be applied. The
consistent definition of mass and angular momenta for the solution \eqref{SL_metric} remains thus an
open question.

If the horizon is not compactified by considering a suitable quotient space of
$\SL(2,\RR)$, one can only define an entropy density, which can be computed by means of the
Bekenstein-Hawking formula and reads
\begin{equation}
s = \frac{S}{V_{\SL(2,\RR)}} = \frac{R_0^3}{32}\sqrt{\frac{3R_0^2}{4\ell^2} - 1}\,,
\end{equation}
where we set Newton's constant $G=1$. Since the black hole \eqref{SL_metric} is extremal, its Hawking
temperature vanishes. The electric charge density is given by
\begin{equation}
q = \frac{Q}{V_{\SL(2,\RR)}} = \frac1{V_{\SL(2,\RR)}}\int\star F\,,
\end{equation}
where the integral is computed on a slice of constant $t,R$, taking subsequently the limit $R\to\infty$.
The result is
\begin{equation}
q = \frac{\sqrt3 R_0^2}8\left(\frac{R_0^2}{2\ell^2} -1\right)\,.
\end{equation}
Note that the norm of the Killing vector $\partial_z$ in \eqref{SL_metric} becomes negative for
$R>R_{\text{VLS}}$, with $V(R_{\text{VLS}})=0$. Using \eqref{constr-R_0}, it is easy to see that
$R_{\text{VLS}}$ lies always outside the horizon, $R_{\text{VLS}}>R_0$. The subspace $R=R_{\text{VLS}}$
was called ``velocity of light surface'' (VLS) in \cite{Gibbons:1999uv} (cf.~also \cite{Caldarelli:2001iq}).
However, contrary to what happens in \cite{Gibbons:1999uv,Caldarelli:2001iq}, here we have no
closed timelike curves (CTCs) for $R>R_{\text{VLS}}$, unless $\SL(2,\RR)$ is compactified to some
quotient space in which $\partial_z$ has closed orbits. Moreover, one easily checks that the
metric \eqref{SL_metric} still keeps the correct signature beyond the VLS.

We close this section with an analysis of the Killing spinors admitted by the solution \eqref{SL_metric},
\eqref{SL_gauge}. The Killing spinor equations are obtained requiring the vanishing of the supersymmetry
variations of the gravitino $\psi_\mu$,
\begin{equation}
\delta\psi_\mu = \biggl[\mathcal{D}_\mu - \frac{i}{4\sqrt3} F_{\nu\rho} (\Gamma_\mu^{\,\ \nu\rho} -
4\delta^\nu_\mu\Gamma^\rho) - \frac1{2\ell}\Gamma_\mu - \frac{i\sqrt3}{\ell} A_\mu\biggr]\kappa\,,
\end{equation}
where $\kappa$ is the supersymmetry parameter and $\mathcal{D}_\mu=\partial_\mu
+\frac14\omega_\mu^{\ ab}\Gamma_{ab}$ denotes the Lorentz-covariant derivative.
$\delta\psi_\mu\equiv\hat{\mathcal{D}}_\mu\kappa = 0$ implies the integrability conditions
\begin{equation}
\hat{\mathcal{R}}_{\mu\nu}\kappa\equiv\bigl[\hat{\mathcal{D}}_\mu,\hat{\mathcal{D}}_\nu\bigr]\kappa
= 0\,, \label{int-cond-KS}
\end{equation}
which admit a nontrivial solution iff $\det(\hat{\mathcal{R}}_{\mu\nu})=0$. This gives a series of
projection conditions to be imposed on $\kappa$. To proceed, we introduce the orthonormal basis
\begin{equation}
\begin{gathered}
e^0 = f (dt + \Psi\sigma_L^3)\,, \qquad e^1 = U^{-1/2} dR\,, \\
e^2 = \frac{R}2\sigma_L^1\,, \qquad e^3 = \frac{R}2\sigma_L^2\,, \qquad e^4 = \frac{R U^{1/2}}{2f}
\sigma_L^3\,.
\end{gathered}
\end{equation}
Condition \eqref{int-cond-KS} implies
\begin{equation}\label{KSE_projection}
	\Gamma_{14}\kappa = -i\epsilon\kappa\,, \qquad\Gamma_{23}\kappa = i\epsilon\kappa\,,
\end{equation}
which means that the $\SL(2,\RR)$-invariant black hole \eqref{SL_metric}, \eqref{SL_gauge} is one quarter
BPS and preserves thus two of the eight (real) supercharges. The Killing spinor reads
\begin{equation}
	\kappa = f^{1/2}\kappa_0\,,
\end{equation}
where $\kappa_0$ is a constant spinor satisfying \eqref{KSE_projection}. If we consider instead the
near-horizon metric \eqref{SL_metric-nh}, the integrability conditions yield the only projection relation
\begin{equation} \label{KSE_projection-nh}
	\Gamma_{23}\kappa = i\epsilon\kappa\,,
\end{equation}
so that we have a supersymmetry enhancement to one half in the near-horizon limit. Again, we can
express $\kappa$ in terms of constant spinors as
\begin{equation}
	\kappa = \zeta^{1/2}\kappa_0^- + \zeta^{-1/2}\kappa_0^+ + \left( i\frac{4W^{1/2}}{R_0} \zeta^{1/2} T - \frac{3R_0 W^{-1/2}}{2\ell} \zeta^{-1/2}\right)\Gamma_1\kappa_0^+\,,
\end{equation}
where $\kappa_0^\pm$ must satisfy \eqref{KSE_projection-nh} and
\begin{equation}
	\Gamma_{14}\kappa_0^\pm = \pm i\epsilon\kappa_0^\pm\,.
\end{equation}

\subsection{Dimensional reduction}

The black hole \eqref{SL_metric}, \eqref{SL_gauge} can be dimensionally reduced to
$D=4$ along the $z$-direction via the $r$-map (see appendix B of \cite{Klemm:2016kxw} for details).
This gives a solution of the t$^3$ model, whose bosonic fields comprise the metric, two gauge fields and a complex scalar. Using the Kaluza-Klein ansatz \cite{Klemm:2016kxw}\footnote{Here $x^\mu = (t,R,x,y)$ are the coordinates on the four-dimensional spacetime.}
\begin{equation} \label{eq:ansatzKK}
	ds_5^2 = e^{\frac{\phi}{\sqrt3}} ds_4^2 + e^{-\frac{2\phi}{\sqrt3}}(dz + K_\mu dx^\mu)^2\,, \quad
	A = B (dz + K_\mu dx^\mu) + C_\mu dx^\mu\,,
\end{equation}
we obtain the four-dimensional metric, the two gauge fields and the complex scalar
\begin{equation}
	\begin{split}
		ds_4^2 &= -\frac{R U}{2\sqrt V} dt^2 + \frac{R\sqrt V}{2U} dR^2 + \frac{R^3\sqrt V}8\frac{dx^2 + dy^2}{x^2}\,, \\
		A^0 &= \frac1{\sqrt2}\biggl(-\frac{4f^2\Psi}{R^2 V} dt + \frac{dy}x\biggr)\,, \qquad A^1 = \frac f{\sqrt2} \biggl(1 + \frac{\epsilon R_0^4 f\Psi}{\ell R^4 V}\biggr) dt\,, \\
		\tau &= -\frac{\epsilon R_0^4}{4\ell R^2} + i\frac{R\sqrt V}2\,.
	\end{split}
\end{equation}
Functions $f(R)$, $\Psi(R)$, $U(R)$ and $V(R)$ are the same as in~\eqref{SL_f+Psi} and~\eqref{SL_U+V}, while $\epsilon=\pm1$.
Like in the five-dimensional solution, we have a regular horizon at $R = R_0$, where $f$, and hence $U$, vanishes. The divergency of the Kretschmann scalar at the locus where $V=0$ indicates the presence of a curvature singularity therein. This turns out to be a naked singularity, since, as already mentioned, $V$ vanishes outside the horizon. Our four-dimensional spacetime is thus pathological.

\section{Matter-coupled case}
\label{sec:matter-coupled}

We shall now generalize the $\SL(2,\RR)$-invariant solution \eqref{SL_metric}, \eqref{SL_gauge}
to the case of $N=2$, $D=5$ U(1)-gauged supergravity coupled to $n$ abelian vector mutiplets. Its bosonic filed content includes the f\"unfbein $e^a_\mu$, the vectors $A^I_\mu$, with $I=1,\ldots,n+1$, and the real scalars $\phi^i$, with $i=1,\ldots,n$. The scalar fields can be conveniently parameterized with $n+1$ functions $h^I=h^I(\phi^i)$ satisfying the condition $\frac16 C_{IJK} h^I h^J h^K = 1$, with $C_{IJK}$ a fully symmetric, constant and real tensor.
The action is~\cite{Gunaydin:1984ak}\footnote{We refer to, e.g., \cite{Klemm:2016kxw} for further details on the theory.}
\begin{equation} \label{IJK_action}
	\begin{split}
		\mathcal{S} &= \frac1{16\pi G} \int \biggl[ (R - 2\mathcal{V})\star\!1 - G_{IJ} \star\!dh^I \wedge dh^J - G_{IJ} \star\!F^I\wedge F^J \\
		& \quad\, - \frac16 C_{IJK} F^I \wedge F^J\wedge A^K \biggr]\,,
	\end{split}
\end{equation}
where $F^I = dA^I$ are the abelian field strengths and the scalar potential reads
\begin{equation}
	\mathcal{V} = 9 g^2 V_I V_J\biggl(\frac12 G^{IJ} - h^I h^J\biggr) \,.
\end{equation}
Here, $g$ is the gauge coupling constant, $V_I$ are Fayet-Iliopoulos parameters and $G^{IJ}$ is the inverse of the kinetic matrix $G_{IJ}$, which can be expressed as
\begin{equation}
	G_{IJ} = \frac92 h_I h_J - \frac12 C_{IJK} h^K \,,
\end{equation}
having defined $h_I \equiv \frac16 C_{IJK} h^J h^K$. Notice that $G_{IJ}$ behaves almost like a metric for the functions $h^I$, indeed $h_I = \frac23 G_{IJ} h^J$.
In what follows, we shall restrict to locally symmetric target spaces parameterized by~$\phi^i$, in which case~\cite{Gunaydin:1984ak}
\begin{equation}
	C_{IJK} C_{J'(LM} C_{PQ)K'} \delta^{JJ'} \delta^{KK'} = \frac43 \delta_{I(L} C_{MPQ)} \,.
\end{equation}
Defining $C^{IJK} \equiv \delta^{II'} \delta^{JJ'} \delta^{KK'} C_{I'J'K'}$, the inverse matrix $G^{IJ}$ reads
\begin{equation}
	G^{IJ} = 2h^I h^J - 6C^{IJK} h_K  \qquad  \implies  \qquad  h^I = \frac92 C^{IJK} h_J h_K \,,
\end{equation}
and the scalar potential can be written compactly as
\begin{equation}
	 \mathcal{V} = -27 g^2 C^{IJK} V_I V_J h_K \,.
\end{equation}

A family of supersymmetric asymptotically AdS$_5$ black holes with spherical horizon was constructed in~\cite{Gutowski:2004yv} through a detailed analysis of the conditions that supersymmetry imposes on the structure of the solution. In this section we shall follow their strategy, applying it to the case of spatial cross-sections modelled on the $\SL(2,\RR)$ geometry.
As a first step we observe that, like in the minimal theory, we can build a scalar $f$, a vector $V$ and three two-forms $J^{(i)}$ out of the Killing spinor. These quantities have to satisfy the same algebraic constraints \eqref{V2-f2}--\eqref{hyperK3} already encountered~\cite{Gutowski:2004yv}.
From the supersymmetry variations it is possible to derive the following set of differential relations~\cite{Gutowski:2004yv}
\begin{subequations}
	\begin{gather}
		df = -i_V (h_I F^I) \,, \label{IJK_df} \\
		\nabla_{(\mu} V_{\nu)} = 0 \,, \\
		dV = -2 f h_I F^I - h_I \star\!(F^I \wedge V) - 2g V_I h^I J^{(1)} \,,  \label{IJK_dV} \\
		dJ^{(i)} = 3g \, \varepsilon_{1ij} V_I \bigl[ A^I \wedge J^{(j)} + h^I \star\!J^{(j)} \bigr] \,.
	\end{gather}
\end{subequations}
Again, $V$ is a Killing vector which leaves the field strengths~$F^I$ invariant~\cite{Gutowski:2004yv}. Moreover, $dJ^{(1)}=0$, i.e.\ $J^{(1)}$ is closed.
Additionally, the dilatino equation implies~\cite{Gutowski:2004yv}
\begin{equation} \label{IJK_F}
	\biggl(\frac14 G_{IJ} - \frac38 h_I h_J\biggr) F^J_{\mu\nu} (J^{(i)})^{\mu\nu} = -\frac{3g}{2} \, \delta^{1i} (h_I V_J h^J - V_I) f \,.
\end{equation}

Like in the minimal theory, we consider $V$ to be not globally null and introduce the timelike ansatz \eqref{timelike-metric}. Also in this case, the base space~$\mathcal{B}$ must be a K\"ahler manifold with anti-selfdual K\"ahler form~$J^{(1)}$ \cite{Gutowski:2004yv}.
Closely following~\cite{Gutowski:2004yv}, to which we refer for further details, from equations~\eqref{IJK_df}, \eqref{IJK_dV} and~\eqref{IJK_F} we obtain the Maxwell field strengths
\begin{equation} \label{IJK_gauge}
	F^I = d[h^I f (dt + \omega)] + \Theta^I - 9 g f^{-1} C^{IJK} V_J h_K J^{(1)} \,,
\end{equation}
where $\Theta^I$ is a selfdual two-form on $\mathcal{B}$ such that
\begin{equation} \label{IJK_theta-1}
	h_I \Theta^I = -\frac23 G^+ \,.
\end{equation}
Like in the minimal theory, we split $d\omega$ into selfdual ($G^+$) and anti-selfdual ($G^-$) parts as in~\eqref{eq:G+-}.
The warp factor $f$ is given by
\begin{equation} \label{IJK_f}
	f = -\frac{108g^2}R C^{IJK} V_I V_J h_K \,,
\end{equation}
with $R$ Ricci scalar of $\mathcal{B}$. Supersymmetry imposes one last constraint,
\begin{equation} \label{IJK_theta-2}
	3g V_I \Theta^I = \mathcal{R} - \frac R4 J^{(1)} \,,
\end{equation}
where $\mathcal{R}$ is the Ricci form defined in~\eqref{ricci-form}.\\
The conditions presented so far are both necessary and sufficient for the existence of a Killing spinor or, put in other words, for our configuration to be supersymmetric. In order to ensure that this background is also a solution of the equations of motion we need to impose the Bianchi identity and the Maxwell equations. Given the field strengths~\eqref{IJK_gauge}, they read, respectively,
\begin{align}
	\label{IJK_bianchi}
	d\Theta^I &= 9 g C^{IJK} V_J \, d\bigl(f^{-1} h_K\bigr) \wedge J^{(1)} \,, \\
	\begin{split} \label{IJK_maxwell}
		d\star_4\! d\bigl(f^{-1} h_I\bigr) &= -\frac16 C_{IJK}\Theta^J\wedge\Theta^K + 2 g f^{-1} V_I G^- \wedge J^{(1)} \\
		& \phantom{{}=} + 6 g^2 f^{-2}\bigl(G_{IJ} C^{JKL} V_K V_L + V_I V_J h^J\bigr) \, \mathrm{vol}_\mathcal{B} \,.
	\end{split}
\end{align}
Once these are solved, the Einstein equations and the equations for the scalars are implied by the Killing spinor equations and are thus satisfied.\\
Here are the ingredients to construct timelike supersymmetric solutions to matter-coupled five-dimensional gauged supergravity.
Pick a K\"ahler space~$\mathcal{B}$ with K\"ahler form~$J^{(1)}$.
Assume $f$ is given by~\eqref{IJK_f} and determine $\omega$, $h^I$ and $\Theta^I$ by means of \eqref{IJK_theta-1}, \eqref{IJK_theta-2}, \eqref{IJK_bianchi} and \eqref{IJK_maxwell}.
The total metric and field strengths are given by~\eqref{timelike-metric} and~\eqref{IJK_gauge}.
The resulting background will be supersymmetric and, in particular, will preserve at least one quarter of the total number of supersymmetries~\cite{Gutowski:2004yv}.
Like in the minimal theory, there is one caveat: not all K\"ahler bases may give rise to a solution. Indeed, all the ingredients must be such that $f^{-1}(G^+ + G^-)$ is a closed two-form, thus having a nontrivial constraint on these quantities.

\subsection{The black hole solution}

Inspired by section~\ref{sec:SL}, we choose for the base space the $\SL(2,\RR)$-invariant ansatz
\begin{equation}
	\begin{split}
		ds_4^2 &= d\rho^2 + a^2\bigl[(\sigma_L^1)^2 + (\sigma_L^2)^2\bigr] + (2a a')^2 (\sigma_L^3)^2\,, \\
		J^{(1)} &= -\epsilon \, d[a^2 \sigma_L^3]\,,
	\end{split}
\end{equation}
with $a=a(\rho)$, $\epsilon=\pm1$ and $\sigma_L^i$ given in \eqref{sl_forms}.
Assuming $a,a'>0$ we introduce the orthonormal frame
\begin{equation}
	e^1 = d\rho\,, \qquad e^2 = a\sigma_L^1\,, \qquad e^3 = a\sigma_L^2\,, \qquad e^4 = 2a a' \sigma_L^3\,,
\end{equation}
in terms of which $J^{(1)}=-\epsilon (e^1\wedge e^4 - e^2\wedge e^3)$.
$f$ is determined by \eqref{IJK_f},
\begin{equation} \label{IJK-eq6}
	f = \frac{54 g^2 a^2 a' C^{IJK} V_I V_J h_K}{4(a')^3 + 7a a' a'' + a^2 a''' + a'}\,.
\end{equation}
The ansatz for the one-form $\omega$ is, again,
\begin{equation}
	\omega = \Psi(\rho)\sigma_L^3\,,
\end{equation}
which, by means of \eqref{eq:G+-}, gives
\begin{equation}
	\begin{split}
		G^+ &= \frac{f a}{4a'}\partial_\rho\bigl(a^{-2}\Psi\bigr) (e^1\wedge e^4 + e^2\wedge e^3)\,, \\
		G^- &= \frac f{4a^3 a'}\partial_\rho\bigl(a^2\Psi\bigr) (e^1\wedge e^4 - e^2\wedge e^3)\,.
	\end{split}
\end{equation}
We assume that the scalars $h^I$ only depend on $\rho$,
\begin{equation}
	h^I = h^I(\rho)\,,
\end{equation}
and, in order to proceed, we adopt the following ansatz for the vector fields
\begin{equation}
	A^I = h^I f (dt + \omega) + U^I(\rho)\sigma_L^3\,.
\end{equation}
Computing $F^I=dA^I$ and comparing with \eqref{IJK_gauge} one finds
\begin{align}
	\label{IJK-Theta^I}
	\Theta^I &= \frac a{4a'}\partial_\rho\bigl(a^{-2} U^I\bigr) (e^1\wedge e^4 + e^2\wedge e^3)\,, \\
	\label{IJK-eq1}
	\partial_\rho\bigl(a^2 U^I\bigr) &= 36\epsilon g f^{-1} a^3 a' C^{IJK} V_J h_K\,.
\end{align}
The last equation ensures that the Bianchi identities \eqref{IJK_bianchi} are automatically satisfied. Contracting \eqref{IJK-Theta^I} with $h_I$ leads to
\begin{equation} \label{IJK-eq2}
	f^{-1} h_I\partial_\rho\bigl(a^{-2} U^I \bigr) = -\frac23\partial_\rho\bigl(a^{-2}\Psi\bigr)\,,
\end{equation}
while a contraction with $V_I$ gives
\begin{equation}
	\frac{3g}{2} V_I\partial_\rho\bigl(a^{-2} U^I \bigr) = -\epsilon\frac{4(a')^3 - 3a a' a'' - a^2 a''' + a'}{a^3}\,,
\end{equation}
which can be integrated, with the result
\begin{equation} \label{IJK-eq3}
	3g V_I U^I = \epsilon\bigl(4(a')^2 + 2a a'' + 1\bigr)\,.
\end{equation}
One can shew that \eqref{IJK-eq1} together with \eqref{IJK-eq3} satisfy equ.~\eqref{IJK-eq6}.
With all these ingredients, the Maxwell equations \eqref{IJK_maxwell} take the form
\begin{equation} \label{IJK-eq4}
	\partial_\rho\biggl[a^3 a'\partial_\rho\bigl(f^{-1} h_I\bigr) + \epsilon g a^2\Psi V_I + \frac1{12} C_{IJK} U^J U^K\biggr] = 0\,.
\end{equation}
In the same spirit of~\cite{Gutowski:2004yv}, we make the following ansatz on the scalar fields:
\begin{equation}
	f^{-1} h_I = \bar X_I + \frac{q_I}{4a^2}\,, \qquad \bar{X}_I = \ell g V_I\,,
\end{equation}
where $q_I$ are free constant parameters. $\ell$ is the AdS$_5$ radius of the minimal theory, which is retrieved from the matter-coupled supergravity when the scalars are fixed to the constant values $h_I = \bar{X}_I$. Since, by definition, $C^{IJK} h_I h_J h_K = \frac29$, we can write
\begin{equation}
	f	= \Bigl(1 + \frac{\alpha_1}{4a^2} + \frac{\alpha_2}{16a^4} + \frac{\alpha_3}{64a^6}\Bigr)^{-1/3}\,,
\end{equation}
having defined the constants
\begin{equation} \label{IJK_alpha}
	\alpha_1 = \frac{27}2 C^{IJK} \bar{X}_I \bar{X}_J q_K\,, \quad \alpha_2 = \frac{27}{2} C^{IJK} \bar{X}_I q_J q_K\,, \quad \alpha_3 = \frac92 C^{IJK} q_I q_J q_K\,.
\end{equation}
Integrating \eqref{IJK-eq1} one gets
\begin{equation}
	U^I = \frac{9\epsilon}{\ell} C^{IJK}\bar{X}_J\Bigl(a^2\bar{X}_K + \frac{q_K}2\Bigr)\,,
\end{equation}
where we ignored an $a^{-2}$ integration term since it is not present in the solution of the minimal theory. Integration of \eqref{IJK-eq2} leads to
\begin{equation} \label{IJK-Psi}
	\Psi = \frac{\epsilon}{\ell}\Bigl(\Psi_0 a^2 - \frac{\alpha_1}2 - \frac{\alpha_2}{16a^2}\Bigr)\,.
\end{equation}
On the other hand, equ.~\eqref{IJK-eq3} yields
\begin{equation} \label{IJK-eq5}
	(a')^2 = \frac{a^2}{\ell^2} + \frac14\Bigl(\frac{\alpha_1}{\ell^2} - 1\Bigr) + \frac{a'_0}{a^4}\,.
\end{equation}
In \eqref{IJK-Psi} and \eqref{IJK-eq5}, $\Psi_0$ and $a'_0$ denote integration constants that are fixed by the Maxwell equations and by comparison with the minimal theory to
\begin{equation}
	\Psi_0 = -2\,, \qquad a'_0 = 0\,.
\end{equation}
\eqref{IJK-eq5} is easily solved, and gives
\begin{equation}
	a = \frac{\ell}2\sqrt{\frac{\alpha_1}{\ell^2} - 1} \, \sinh(\rho/\ell)\,.
\end{equation}

In order to write down the solution in a simple fashion, we define the new radial coordinate\footnote{Notice
that this definition differs from the one used in the minimal theory by a factor of $f$.}
\begin{equation}
R = 2a\,.
\end{equation}
With this choice, the metric, gauge potentials and scalar fields are given by
\begin{equation}
	\begin{split}
		ds^2 &= -f^2 (dt + \Psi \sigma_L^3)^2 + \frac{dR^2}U + \frac{R^2}4 f^{-1}\bigl[(\sigma_L^1)^2 + (\sigma_L^2)^2 + f^{-1} U (\sigma_L^3)^2\bigr]\,, \\
		A^I &= h^I f dt + \bigl(h^I f\Psi + U^I\bigr)\sigma_L^3\,, \\
		h_I &= f \biggl(\bar{X}_I + \frac{q_I}{R^2}\biggr)\,,
	\end{split}
\end{equation}
with the functions
\begin{equation}
	\begin{aligned}
		f(R) &= \biggl(1 + \frac{\alpha_1}{R^2} + \frac{\alpha_2}{R^4} + \frac{\alpha_3}{R^6}\biggr)^{-1/3}\,, \quad
		& \Psi(R) & = -\frac{\epsilon R^2}{2\ell} \biggl(1 + \frac{\alpha_1}{R^2} + \frac{\alpha_2}{2R^4}\biggr)\,, \\
		U(R) &= f(R) \biggl(-1 + \frac{\alpha_1}{\ell^2} + \frac{R^2}{\ell^2}\biggr)\,, \quad
		& U^I &= \frac{9\epsilon}{4\ell} C^{IJK}\bar{X}_J\bigl(\bar{X}_K R^2 + 2q_K\bigr)\,.
	\end{aligned}
\end{equation}
We recall that $\bar{X}_I = \ell g V_I$, while the constants $\alpha_i$ are defined in~\eqref{IJK_alpha}.
The solution of the minimal theory is retrieved setting $\alpha_1=3R_0^2$, $\alpha_2=3R_0^4$ and $\alpha_3=R_0^6$ and redefining the radial coordinate as $R^2\mapsto R^2-R_0^2$. The gauge potential is given by the linear combination $F=\frac{\sqrt3}{2}h_I F^I$.

\subsection{stu model}

One of the most studied examples of matter-coupled theories is the stu model. This system comprises two vector multiplets, whose scalar fields can be parameterized by three real functions such that $h^1 h^2 h^3=1$, thus, in the notation of this section, with only nonvanishing component $C_{123}=1$ (and permutations).
This model can be embedded in higher-dimensional supergravity theories, arising as consistent truncation of type-IIB supergravity on S$^5$ or M-theory compactified on a Calabi-Yau.\\
In the stu model we have
\begin{equation}
	h_I = \frac13 (h^I)^{-1}\,, \qquad G_{IJ} = \frac92\diag(h_1^2, h_2^2, h_3^2)\,.
\end{equation}
For simplicity, we also set $gV_I=1/(3\ell)$, hence $\bar{X}_I=1/3$.
In this case the solution simplifies drastically. Indeed, defining the functions
\begin{equation}
	H_I = 1 + \frac{\mu_I}{R^2}\,,
\end{equation}
with $\mu_I=3q_I$, one can write $f$ as
\begin{equation}
	f = (H_1 H_2 H_3)^{-1/3}\,,
\end{equation}
whereas the gauge potentials and the scalar fields become
\begin{equation}
	\begin{split}
		A^I &= H_I^{-1} dt + \frac{\epsilon}{4\ell R^2} \biggl(\alpha_2 - \frac{2\alpha_3}{\mu_I}\biggr) H_I^{-1}\sigma_L^3\,, \\
		h^I &= (H_1 H_2 H_3)^{1/3} H_I^{-1}\,.
	\end{split}
\end{equation}
The constants $\alpha_I$ are related to the charges $\mu_I$ by
\begin{equation}
	\alpha_1 = \mu_1 + \mu_2 + \mu_3\,, \qquad \alpha_2 = \mu_1\mu_2 + \mu_2\mu_3 + \mu_3\mu_1\,, \qquad \alpha_3 = \mu_1\mu_2\mu_3\,.
\end{equation}
The solution of the minimal theory corresponds to $\mu_I=R_0^2$, together with the change of the
radial coordinate $R^2\mapsto R^2-R_0^2$.

\section{Final remarks}
\label{sec:final-rem}

We conclude this paper with a list of possible future extensions. First of all, it would be
interesting to construct rotating, charged (and supersymmetric) black holes with homogeneous
horizons also in dimension $D>5$.

In all cases considered in this paper, the horizon geometry was either Nil or $\SL(2,\RR$), but
probably also the solutions of \cite{Cadeau:2000tj,Faedo:2019rgo}, whose spatial cross-sections
are foliated by three-manifolds Sol, should admit rotating generalizations. The results
of~\cite{Gutowski:2004ez} strongly suggest that these have no BPS limit, since the group manifold
Sol does not appear in the possible near-horizon geometries\footnote{In the static case, it was
explicitely shown in \cite{Faedo:2019rgo} that there are no supersymmetric Sol-invariant black
holes.}. Furthermore, ref.~\cite{Grover:2013hja} implies that no supersymmetric limit can exist when a compact quotient of the solvegeometry horizon is considered.

In this context, it is also amusing to note that ref.~\cite{Coiculescu:2020xx} studies a one-parameter
family of nonisomorphic solvable Lie groups, which, when equipped with canonical left-invariant metrics
\begin{equation} \label{metr-interpolation}
	ds^2 = e^{-2z} dx^2 + e^{2\alpha z} dy^2 + dz^2\,,
\end{equation}
interpolates from Sol geometry ($\alpha=1$) to hyperbolic space $\text{H}^3$
($\alpha=-1$), with an intermediate stop at $\text{H}^2\times\RR$ for $\alpha=0$.
Static black holes with Sol and $\text{H}^3$ horizons were found in \cite{Cadeau:2000tj,Faedo:2019rgo} and \cite{Birmingham:1998nr} respectively, while the case $\alpha=0$ corresponds to the black strings of \cite{Klemm:2000nj,Bernamonti:2007bu}. However, the existence of solutions with generic horizon geometry \eqref{metr-interpolation} remains an open question.

Recently, in~\cite{Lucietti:2021bbh,Lucietti:2022fqj} a classification of supersymmetric black holes in minimal gauged supergravity that admit an SU(2) or a U(1)$^2$ isometry was performed. Starting from these results, unique theorems were then proven for AdS$_5$ black holes with spherical horizon or toric symmetry.
Following a similar strategy, an extension of these classifications could be attempted, considering Nil and $\SL(2,\RR)$ as symmetry groups.

A further point concerns the static Nil black hole of \cite{Cadeau:2000tj}. We conjecture that, similar to \eqref{metr-Nil-nonextr-rot}, it arises as a scaling limit of a static anisotropic spherical black hole yet to be constructed.

Lastly, as we already mentioned, we expect that novel nonextremal Nil black holes could be obtained applying the scaling limit of subsection~\ref{subsec:Nonextr-Nil} to the spherical, charged and rotating solutions of~\cite{Chong:2005hr}.
In a similar way, new solutions to the matter-coupled theory could arise employing an analogous prescription to the supersymmetric multi-charge black holes constructed in~\cite{Kunduri:2006ek}.

We hope to come back to these points in a forthcoming publication.

\section*{Acknowledgements}

This work was supported partly by INFN.
The research of FF is supported by the project ``Nonlinear Differential Equations'' of Universit\`a degli Studi di Torino.

\appendix

\section{Homogeneous manifolds}
\label{app:homogeneous-spaces}

Let $(M, g)$ be a (pseudo-)Riemannian manifold with isometry group $G$. $M$ is said to be homogeneous if $G$ acts transitively on $M$, i.e., if $\forall\ p,q\in M$ there exists an isometry $\phi\in G$ such that $\phi(p)=q$. If the element $\phi$ is unique we say that $G$ acts simply transitively on $M$; this is equivalent to saying that $\dim G = \dim M$. In this case, $M$ is said to be simply transitive too.\\
For simply transitive manifolds, since $\dim G = \dim M$, the Killing vectors $\xi_A$ ($A=1,\dots,\dim M$) form a basis of the tangent space. However, it is possible to construct a $G$-invariant basis $X_A$ \cite{Ryan:1975jw} such that
\begin{equation}
	\mathcal{L}_{\xi_B} X_A = [\xi_B, X_A] = 0\quad\forall\, A, B\,,
\end{equation}
where $\mathcal{L}_{\xi_B} X_A$ is the Lie derivative of $X_A$ along the direction of $\xi_B$.
The description of a homogeneous manifold in terms of an invariant basis is particularly simple since the components of the metric $g_{AB}$ are group invariant, and hence constant, and it is possible to write the metric in the form
\begin{equation}
	ds^2 = g_{AB} \sigma_L^A \sigma_L^B\,,
\end{equation}
where the $\sigma_L^A$ form the dual basis of $X_A$. The dual basis is $G$-invariant as well, $\mathcal{L}_{\xi_B}\sigma_L^A=0$, and satisfies
\begin{equation}
	d\sigma_L^A = \frac12 C^A_{\,\ BC} \sigma_L^B \wedge \sigma_L^C\,,
\end{equation}
with $C^A_{\,\ BC}$ the structure constants of the Lie algebra of $G$.\\
There are nine three-dimensional Lie algebras, the Bianchi cosmologies, labelled from type I to type IX.
The Bianchi cosmologies are in correspondence with the eight Thurston model geometries, that play an essential role in the Thurston conjecture~\cite{Thurston:1997}.\\
Below we list the metrics of nilgeometry (type II) and $\SL(2,\RR)$ (type VIII) in terms of the $G$-invariant dual basis $\sigma_L^A$ together with the nonvanishing structure constants of the corresponding Lie algebras. We use the same one-forms and structure constants that are used in \cite{Gutowski:2004ez} in the near-horizon analysis.

\begin{itemize}
\item Nilgeometry
\begin{subequations}
\begin{gather}
C^3_{\,\ 12} = -C^3_{\,\ 21} = -1\,, \\
\label{nil_forms}
\sigma_L^1 = dx\,, \qquad \sigma_L^2 = dy\,, \qquad \sigma_L^3 = dz + \frac{y dx - x dy}{2}\,, \\
ds^2 = dx^2 + dy^2 + \biggl(dz + \frac{y dx - x dy}{2}\biggr)^2\,,
\end{gather}
\end{subequations}

\item $\SL(2,\RR)$
\begin{subequations}
\begin{gather}
C^1_{\,\ 23} = -C^1_{\,\ 32} = 1\,, \quad C^2_{\,\ 31} = -C^2_{\,\ 13} = 1\,, \quad C^3_{\,\ 12} =
-C^3_{\,\ 21} = -1\,, \\
\label{sl_forms}
\sigma_L^1 + i\sigma_L^2 = \frac{e^{iz}}x (dx + i dy)\,, \qquad \sigma_L^3 = dz + \frac{dy}x\,, \\
ds^2 = \frac{dx^2 + dy^2}{x^2} + \biggl(dz + \frac{dy}{x}\biggr)^2\,.
\end{gather}
\end{subequations}
\end{itemize}

\end{document}